\documentclass[11pt]{article}
\usepackage{graphicx}
\usepackage{amsmath}
\usepackage{amssymb}
\usepackage{amsxtra}
\usepackage{euscript}
\usepackage{slashed}

\newcommand{\be}{\begin{eqnarray}}
\newcommand{\ee}{\end{eqnarray}}
\newcommand{\0}{\nonumber}
\newcommand{\td}{\tilde\partial}

\newcommand{\arccot}{{\rm arcot}}
\newcommand{\EW}{\EuScript W}

\begin{document}
\title{Pure contact term correlators in CFT}
\author{Loriano Bonora\\International School for Advanced Studies (SISSA),\\Via
Bonomea
265, 34136 Trieste and INFN, Sezione di Trieste, Italy,\\ bonora@sissa.it\\ \\Bruno
Lima de Souza\\ International School for Advanced Studies (SISSA),\\Via Bonomea
265, 34136 Trieste and INFN, Sezione di Trieste, Italy, \\blima@sissa.it}

\maketitle

\begin{abstract}
We discuss the case of correlators in CFT made of pure contact terms, without a corresponding bare part.
We show two examples. The first is provided by the conformal limits of a free 
massive fermion theory in 3d. We show that the (conserved) current correlators are in one-to-one 
correspondence with the terms of the 3d  gauge CS action.
The second is the Pontryagin trace anomaly. The corresponding 3-point correlator 
is nonvanishing even though the corresponding untraced 
correlator vanishes.
\end{abstract}
\section{Introduction}\label{s:intro}

Correlators in conformal field theories can be formulated
both in configuration space and, via Fourier transform, in momentum space. In
the first form they may happen to be singular at coincident insertion points and
in need to be regularized. In coordinate space they are therefore simply
distributions. In the simplest cases such distributions have been studied and
can be found in textbooks. But in general the 
correlators of CFT are very complicate expressions and their regularization has to
be carried out from scratch.
This can be done directly in configuration space, in which case a well known
procedure is the differential regularization. An alternative, and often more
accessible, technique consists in formulating the same problem in momentum space
via Fourier transform and proceeding to regularize the Fourier transform of the
relevant
correlators. This procedure produces various types of terms, which we refer to as
{\it non-local, partially local and local terms}. Local terms are represented
by polynomials of the external momenta in momentum space,  and by delta
functions and derivatives of delta functions in configuration space. The
unregularized correlators will be referred to as {\it bare} correlators; they are 
ordinary regular functions at non-coincident points and are
classified as non-local in the previous classification. While regularizing the latter
one usually produces not only local terms, but also intermediate ones, which are
product of bare functions and delta functions or derivatives thereof. These are
referred to as partially local. 

From the above introduction one might be led to think that local terms (i.e. polynomials of the external 
momenta, in momentum space representation) can come only from regularizing bare
correlators. This is not the case, there are important cases of local
correlators that do not have a bare counterpart. We can say that they consist
only of the quantum part. This is the main subject of this article. We will
discuss two examples. The first, in 3d, is the case of pure contact terms in the parity-odd sector of 
the 2-point function of currents. There exist no bare terms corresponding to them. An important implication 
of these contact terms is that they give rise to a Chern-Simons term in the effective action.

The second example is that of the 3-point function of the energy-momentum tensor in 4d, in which one
of the entries is the trace of the em tensor. Classically, the trace of the em tensor is zero in a Weyl 
invariant theory. At the quantum level this fact becomes a set of Ward identities that relate $n$-point 
functions with one insertion of the trace of the em tensor with $(n-1)$-point functions. When the theory 
possesses trace anomalies these Ward identities are complemented by a set of contact terms which 
reproduces the anomalies. What we would like to stress here is that such correlators containing 
one trace insertion can be nonvanishing even if  there is no bare correlator corresponding to it. 
This is what happens with the Pontryagin trace anomaly. The latter is puzzling at first, but, in 
fact, when properly understood, it would be surprising if it did not exist. 

There are of course other examples, beside the two above ones. All these examples are
characterized by the fact that they break parity.

The paper is organized as follows. In the next section we introduce some basic CFT 
formulas in momentum space. In section 3 we work out the 3d example of pure contact term 
correlators and its connection with gauge CS. In section 4 we review the 4d example, 
which corresponds to the Pontryagin trace anomaly. In section 5 we add some new remarks 
concerning this anomaly.

\section{Conformal invariance in momentum space}\label{s:momentum}

In this section we will lay down some introductory material on conformal
invariance and conformal field theories, which will be needed in the sequel. The
conformal group in $d$ dimension encompasses the Poincar\'e transformations, the
dilatation and the special conformal transformations (SCTs). The latter is 
\be 
x'^{\mu} = \frac {x^\mu+b^\mu x^2}{1+2b\!\cdot \!x+ b^2 x^2} = x^\mu+b^\mu
x^2 - 2b\!\cdot \!x\,x^\mu +\mathcal{O}(b^2) \0
\ee
for infinitesimal $b^\mu$. In this paper we will mostly consider the effects 
of conformal invariance in momentum space.
If we Fourier transform the generators of the conformal algebra we get (a tilde
represents the transformed generator and $\tilde \partial= \frac
{\partial}{\partial k}$)
\be
&&\tilde P_\mu = -k_\mu,\0\\
&& \tilde D = i(d+k^\mu {\tilde\partial}_\mu),\0\\
&& \tilde L_{\mu\nu}= i(k_\mu \td_\nu -k_\nu \td_\mu),\0\\
&&\tilde K_\mu= 2d\ \td_\mu +2 k_\nu \td^\nu \td_\mu - k_\mu \tilde \square. \0
\ee  
Notice that $\tilde P_\mu$ is a multiplication operator and $\tilde K_\mu$ is a
quadratic differential operator. The  Leibniz rule does not 
hold for $\tilde K_\mu$ and $\tilde P_\mu$ with respect to the ordinary product.
However it does hold for the convolution product:
\be 
\tilde K_\mu (\tilde f \star \tilde g) = (\tilde K_\mu \tilde f) \star \tilde g+
  \tilde f \star (\tilde K_\mu\ \tilde g)\0
\ee
where $(\tilde f \star \tilde g)(k)=\int dp \, f(k-p)g(p)$. Nevertheless these generators form a closed algebra 
\be
&& [\tilde D, \tilde P_\mu]=i\tilde P_\mu,\0\\
&& [\tilde D, \tilde K_\mu]=i\tilde K_\mu,\0\\
&& [\tilde K_\mu, \tilde K_\nu]=0,\0\\
&& [\tilde K_\mu,  \tilde P_\nu]= 2i( \eta_{\mu\nu} \tilde D -\tilde L_{\mu\nu}),\0\\
&& [\tilde K_\lambda, \tilde L_{\mu\nu}]=i(\eta_{\lambda\mu}\tilde K_\nu-\eta_{\lambda\nu}\tilde K_\mu),\0\\
&& [\tilde  P_\lambda, \tilde L_{\mu\nu}]=i(\eta_{\lambda\mu}\tilde P_\nu-\eta_{\lambda\nu}\tilde P_\mu),\0\\ 
&&[\tilde L_{\mu\nu},\tilde L_{\lambda\rho}]= i(\eta_{\nu\lambda} \tilde L_{\mu\rho}+ \eta_{\mu\rho} \tilde L_{\nu\lambda} - \eta_{\mu\lambda} \tilde L_{\nu\rho}- \eta_{\nu\rho} \tilde L_{\mu\lambda}.\nonumber
\ee
One should be aware that they do not generate infinitesimal transformation in momentum
space. This notwithstanding, in momentum space we can write down the conformal Ward identities that the correlators must satisfy, 
see \cite{Bzowski:2013sza}. As an example, let us consider the SCT for the 2-point function of a current $J_\mu$ 
and the energy-momentum tensor $T_{\mu\nu}$ in $d$ dimensions. For the 2-point function of currents we have the special conformal Ward identity
\begin{multline}
\mathcal{K}_{\kappa}\langle J_{\mu}(\boldsymbol{k})J_{\nu}(-\boldsymbol{k}) \rangle = (2(\Delta-d) \tilde \partial_\kappa-2k\!\cdot\!\tilde \partial \, \tilde \partial_\kappa +k_\kappa \tilde
\square ) \langle J_{\mu}(\boldsymbol{k})J_{\nu}(-\boldsymbol{k}) \rangle \\+
2(\eta_{\kappa\mu}\tilde{\partial}^\alpha-\delta_{\kappa}^{\alpha}\tilde{
\partial}_\mu)\langle J_{\alpha}(\boldsymbol{k})J_{\nu}(-\boldsymbol{k}) \rangle
= 0,\label{tildeJtransfJ}
\end{multline}
while for the 2-point function of the energy-momentum tensor we have
\begin{multline}
\mathcal{K}_{\kappa} \langle T_{\mu\nu}(\boldsymbol{k})J_{\rho\sigma}(-\boldsymbol{k}) \rangle = (2(\Delta-d) \tilde \partial_\kappa-2k\!\cdot\!\tilde \partial \, \tilde \partial_\kappa +k_\kappa \tilde
\square ) \langle T_{\mu\nu}(\boldsymbol{k})T_{\rho\sigma}(-\boldsymbol{k}) \rangle \\+ 2(\eta_{\kappa\mu}\tilde{\partial}^\alpha-\delta_{\kappa}^{\alpha}\tilde{\partial}_\mu)\langle T_{\alpha\nu}(\boldsymbol{k})T_{\rho\sigma}(-\boldsymbol{k}) \rangle + 2(\eta_{\kappa\nu}\tilde{\partial}^\alpha-\delta_{\kappa}^{\alpha}\tilde{\partial}_\nu)\langle T_{\mu\alpha}(\boldsymbol{k})T_{\rho\sigma}(-\boldsymbol{k}) \rangle = 0.\label{tildeJtransfT}
\end{multline}

\section{2- and 3-point functions and CS effective action}\label{s:CS}

The first example announced in the introduction is mostly pedagogical. It arises from a very simple model,
a free massive fermion model in 3d coupled to a gauge field, see \cite{Babu:1987rs,Dunne:1998qy,Gama:2015tuy}. The action is 
\be
S=\int
d^{3}x\,\left(i\bar{\psi} \gamma^{\mu}D_{\mu}\psi-m\bar{\psi}
\psi\right),\quad D_{\mu}=
\partial_{\mu}+ A_\mu \label{freeaction}
\ee
where $A_\mu= A_\mu^a(x) T^a$ and $T^a$ are the generators of a gauge algebra in
a given representation determined by $\psi$. The generators are antihermitean, 
$[T^a,T^b]=f^{abc} T^c$, with normalization ${\rm tr}(T^a T^b)={\rm
n}\,\delta^{ab}$. 

The current
\be
J^a_\mu(x) = \bar \psi \gamma_\mu T^a \psi\label{Jmu}
\ee
is (classically) covariantly conserved on shell 
\be
(DJ)^a = (\partial^\mu \delta^{ac} + f^{abc} A^{b\mu})
J_\mu^c=0\label{currentconserv}
\ee

The generating functional of the connected Green functions is given by
\be 
W[A]=\sum_{n=1}^\infty \frac {i^{n+1}}{n!} \int \prod_{i=1}^n d^3x_i A^{a_1
\mu_1}(x_1)  \ldots A^{a_n \mu_n}(x_n) 
\langle 0|{\cal T} J_{\mu_1}^{a_1}(x_1)\ldots
J_{\mu_n}^{a_n}(x_n)|0\rangle\label{WA}
\ee
The full 1-point function of $J_\mu^a$ in the presence of the source $A^{a\mu}$ is
\begin{multline}
\langle\!\langle J_\mu^a(x)\rangle\!\rangle = \frac {\delta W[A]}{\delta
A^{a\mu}(x)} = -\sum_{n=1}^\infty \frac {i^n}{n!} \int \prod_{i=1}^n d^3x_i A^{a_1
\mu_1}(x_1)  \ldots A^{a_n \mu_n}(x_n) \\ \langle 0|{\cal T}J_\mu^a(x)
J_{\mu_1}^{a_1}(x_1)
\ldots J_{\mu_n}^{a_n}(x_n)|0\rangle \label{Jamux}
\end{multline}
The 1-loop conservation is
\be
(D_\mu \langle\!\langle J_\mu(x)\rangle\!\rangle)^a= \partial^\mu
\langle\!\langle
J_\mu^a(x)\rangle\!\rangle 
+ f^{abc} A_\mu^b(x) \langle\!\langle J^{\mu
c}(x)\rangle\!\rangle=0\label{qconsrv}
\ee
if there are no anomalies. By deriving this relation with respect to $A$ we find the implications of conservation for the 2-point and 3-point correlators
\begin{gather} 
k^\mu \tilde J_{\mu\nu}^{ab}(k)=0\label{2ptconserv}\\
-i q^\mu \tilde J_{\mu\nu\lambda}^{abc}(k_1,k_2) +f^{abd} \tilde
J_{\nu\lambda}^{dc}(k_2)+ 
f^{acd} \tilde J_{\lambda\nu}^{db}(k_1)=0 \label{3ptconserv}
\end{gather}
where $q=k_1+k_2$ and $\tilde J_{\mu\nu}^{ab}(k)$ and $\tilde
J_{\mu\nu\lambda}^{abc}(k_1,k_2)$ are Fourier transform of the 2- and 3-point functions,
respectively.

The Feynman rules are easily extracted from the action. The propagator is $\frac
i{\slashed {p}-m}$ and the gauge-fermion-fermion vertex is
simply $ \gamma_\mu T^a$, where $\mu, a$ are the labels of
$A_\mu^a$
Our next task will be to calculate the odd-parity
2- and 3-point correlators in this model and study their behavior in the IR and UV limit.

\subsection{The 2-point current correlator}\label{ss:CS2pt}

The relevant diagram is the bubble one, with external momentum $k$. Its Fourier
transform is
\be
&&\tilde J_{\mu\nu}^{ab}(k)=- \int \frac{d^3p}{(2\pi)^3} {\rm Tr}
\left(\gamma_\mu T^a \frac 1{\slashed{p} -m }\gamma_\nu T^b  
\frac 1{\slashed{p}-\slashed{k} -m }\right) = -2{\rm n}
\,\delta^{ab}\label{Jmunuab}\\
&&\cdot \int\frac{d^3p}{(2\pi)^3}\frac {p_\nu(p-k)_\mu
-p\!\cdot\!(p-k)\eta_{\mu\nu} + p_\mu(p-k)_\nu+im \epsilon_{\mu\nu\sigma}
k^\sigma +m^2 \eta_{\mu\nu}}{(p^2-m^2)((p-k)^2-m^2)}\0
\ee 
Let us focus from now on on the odd-parity part. After a Wick rotation and
integration we get
\be
\tilde J_{\mu\nu }^{ab(odd)}(k)&=& \frac {\rm n}{2\pi} \delta^{ab}
\epsilon_{\mu\nu\sigma} k^\sigma \, \frac mk \arctan \frac k{2m}\label{Jmunuodd}
\ee
where $k= \sqrt{k^2}$. The conservation law (\ref{2ptconserv}) is readily seen
to be satisfied.

We are interested in the IR and UV limits of this expression. To this end we
notice that $k$ is the total energy $E$ of the process. Therefore the IR and UV
limit correspond to $\frac m E\to \infty$ and 0, respectively. Therefore near
the IR (\ref{Jmunuodd}) becomes
\be
\tilde J_{\mu\nu }^{ab(odd)}(k)=  \frac {\rm n}{2\pi} \delta^{ab}
\epsilon_{\mu\nu\sigma} k^\sigma\left( \frac 12 -\frac 1{24} \left(\frac k
m\right)^2 +
\frac 1{160} \left(\frac k m\right)^4+\ldots\right)\label{JmnIR}
\ee
and near the UV
\be 
 \tilde J_{\mu\nu }^{ab(odd)}(k)=  \frac {\rm n}{2\pi} \delta^{ab}
\epsilon_{\mu\nu\sigma} k^\sigma\left(  \frac {\pi}2 \frac mk 
- 2 \left(\frac mk\right)^2+ \frac 83 \left(\frac
mk\right)^4+\ldots\right)\label{JmnUV}
\ee
In particular in the two limits we have
\be
\tilde J_{\mu\nu }^{ab(odd)}(k)= \frac {\rm n}{2\pi} \delta^{ab}
\epsilon_{\mu\nu\sigma} k^\sigma \left\{\begin{matrix} \frac 12 & \quad {\rm
IR}\\                                                                           
 
                    \frac {\pi}2 \frac mk & \quad {\rm UV}
                                                                             
                   \end{matrix}\right.\label{JmunuIRUV}
\ee
So far we have worked with a Euclidean metric, but the same result hold also 
for Lorentzian metric (see \cite{next}).

We notice that the UV limit is actually vanishing. However we could consider a
model made of $N$ identical copies of free fermions coupled to the same gauge
field. Then the result (\ref{JmunuIRUV}) would be
\be
\tilde J_{\mu\nu }^{ab(odd)}(k)= \frac {\rm n N}4 \delta^{ab}
\epsilon_{\mu\nu\sigma} k^\sigma  \frac mk \label{JmunuUV}
\ee
In this case we can consider the scaling limit $ \frac mk\to 0$ and $N\to
\infty$ in such a way that $N  \frac mk$ is fixed. Then the UV limit (\ref
{JmunuUV}) becomes nonvanishing.

Before discussing the implications of the previous results let us consider also
the 3-current correlator.

\subsection{The 3-point current correlator}\label{ss:3pt}

The 3-point correlator for currents is given by the triangle diagram. The
three external
momenta are $q,k_1,k_2$. $q$ is incoming, 
while $k_1,k_2$ are outgoing and, of course, momentum conservation implies
$q=k_1+k_2$. The Fourier transform is
\be
\tilde J_{\mu\nu\lambda}^{1,abc}(k_1,k_2)\label{Jmnl}
 =i \int \frac{d^3p}{(2\pi)^3} {\rm
Tr}\left(\gamma_\mu T^a \frac 1{\slashed{p} -m }   \gamma_\nu T^b
\frac 1{\slashed{p}-\slashed{k}_1 -m }\gamma_\lambda T^c\frac
1{\slashed{p}-\slashed{q} -m }\right)
\ee
to which we have to add the cross graph corresponding to the exchange
$b\leftrightarrow c, \nu\leftrightarrow\lambda, 1\leftrightarrow 2$.

We will not go through all the calculation, which is rather more complicated
than in
2-point case. For instance, near the IR fixed point we obtain a series expansion of
the type
\be 
\tilde J_{\mu\nu\lambda}^{1,abc(odd)}(k_1,k_2) \approx i\frac {\rm
n}{32\pi} 
\sum_{n=0}^\infty \left(\frac {E}{m}\right)^{2n} f^{abc} \tilde
I_{\mu\nu\lambda}^{(2n)}(k_1,k_2)\label{IRodd1}
\ee
and, in particular, 
\be
I_{\mu\nu\lambda}^{(0)}(k_1,k_2)= -12 \epsilon_{\mu\nu\lambda}\label{IRodd2}
\ee

Let us pause to comment on this result.
We expect the current (\ref{Jmu}) to be conserved also at the quantum level, because
no anomaly is expected in this case. This should be true
also in the IR  limit. It would seem that conservation, if any, should hold order
by order in the expansions we have considered in (\ref{IRodd1}).  In order to
check conservation we have to verify (\ref{3ptconserv}). Conservation has a
contribution from the 2-point function, so the LHS of equation (\ref{3ptconserv})
reads
\be 
-\frac 3{8\pi} {\rm n} f^{abc} q^\mu \epsilon_{\mu\nu\lambda} + \frac 1{4\pi} 
f^{abc}  \epsilon_{\nu\lambda\sigma}k_2^\sigma +
 \frac 1{4\pi}  f^{abc}  \epsilon_{\nu\lambda\sigma}k_1^\sigma\neq
0.\label{noconserv1}
\ee
Conservation is violated unless we add to $I_{\mu\nu\lambda}^{(0)}(k_1,k_2)$ a
term $4  \epsilon_{\mu\nu\lambda}$. In order to understand what is at stake
here let us turn to the Chern-Simons action for the gauge field $A$ in 3d.

\subsection{The CS action}\label{ss:CSaction}
 
The CS action for the gauge field $A$ is
\be
CS &=& \frac {\kappa}{4\pi} \int d^3x {\rm Tr} \left( A\wedge dA + \frac 23
A\wedge A \wedge A\right)\label{gaugeCS}\\
&=& \frac {{\rm n} \kappa}{4\pi} \int d^3x \epsilon^{\mu\nu\lambda} \left(
A^a_\mu \partial_\nu A^a_\lambda 
+\frac 13 f^{abc} A_\mu^a A_\nu^b A_\lambda^c\right)\0
\ee
  
Now let us return to the 2- and 3-point functions obtained above.
The Fourier anti-transform of the 2-point function $\sim  \epsilon_{\mu\nu\sigma}
k^\sigma $ is
\be
{\cal F}^{-1} [\epsilon_{\mu\nu\sigma} k^\sigma](x)
=i\epsilon_{\mu\nu\sigma}\partial^\sigma \delta(x)\label{Ftransf1}
\ee
The Fourier anti-transform of the 3-point function $\sim  \epsilon_{\mu\nu\lambda}$
is
\begin{multline}
{\cal F}^{-1} [\epsilon_{\mu\nu\sigma}](x,y,z) \label{Ftransf2}\\
= \int \frac
{d^3q}{(2\pi)^3}e^{-iqx}  \int \frac {d^3k_1}{(2\pi)^3} e^{-ik_1 y} \int \frac
{d^3k_2}{(2\pi)^3} 
e^{ik_2 z}\delta(q-k_1-k_2)  \epsilon_{\mu\nu\lambda}\\
=   \int \frac {d^3k_1}{(2\pi)^3} \int \frac {d^3k_2}{(2\pi)^3} e^{ik_1 (y-x)}
 e^{ik_2 (y-z)} \epsilon_{\mu\nu\lambda}=
\delta(y-x) \delta(z-x) \epsilon_{\mu\nu\lambda}
\end{multline}
Inserting this into the functional generator $W[A]$ and integrating with respect
to space time we obtain the two terms of the action
(\ref{gaugeCS}). Therefore if we add to $I_{\mu\nu\lambda}^{(0)}(k_1,k_2)$ a
term $4  \epsilon_{\mu\nu\lambda}$ the effective action
of our model in the IR gives back the CS action with coupling $\kappa=\frac 12$.
 
This corresponds to correcting the effective action by adding a counterterm 
\be
4 \int dx  \epsilon^{\mu\nu\lambda} f^{abc}A_\mu^a A_\nu^b
A_\lambda^c\label{counterIR1}
\ee
This counterterm simultaneously guarantees conservation, see (\ref{noconserv1}),
and reconstructs the correct CS action. We remark that for the effective action induced by a couple of
Majorana fermions, in the IR limit the CS coupling $\kappa=1$, see (\ref{JmunuIRUV}). This
guarantees invariance of the action also under large gauge transformations, \cite{Closset}.

Something similar can be done also for the UV limit. However in the UV limit the
resulting effective action has a vanishing coupling, unless we consider an $N\to
\infty$ limit theory, as outlined above. In order to guarantee invariance under
large gauge transformations we have also to fine tune the limit in such a way
that the $\kappa$ coupling be an integer.

Free fermions in 3d can be coupled also to a background metric. In this case the
relevant correlators are those of the energy-momentum tensor and the resulting
effective action in the UV and IR is the gravitational CS action, see \cite{next}.

\subsubsection{A few remarks}\label{ss:remarks}

We would like to stress a few points of the above construction. The first is the
problem of non-conservation for the 3-point function we have met. This is a consequence of the particular
regularization procedure we have used, that is of the fact the we have first
computed the 3-point function of three currents and then contracted
the correlator with the external momentum $q^\mu$. We could have proceeded in
another way, that is we could have contracted the 3-point correlator with $q^\mu = k_1^\mu+k_2^\mu$
before doing the integration over $p$. The triangle diagram contracted with
$q^\mu$ is:
\be 
q^\mu {\tilde J}^{abc}_{\mu\nu\lambda} (k_1,k_2)  = -i \!\int\! \frac {d^3p}{(2\pi)^3}
{\rm Tr} \left( \slashed{q} T^a \frac 1{\slashed{p}-m} \gamma_\nu T^b \frac
1{\slashed{p}-\slashed{k}_1-m} 
\gamma_\lambda T^c \frac 1{\slashed{p}-\slashed{q}-m} \right).\label{qJ1}
\ee 
Replacing $\slashed{q} =( \slashed{p}-m)- ( \slashed{p}-\slashed{q}-m)$
considerably simplifies the 
calculation. The final result for the odd parity part (after adding the cross
diagram contribution, $1\leftrightarrow 2, b\rightarrow c, \nu\leftrightarrow
\lambda$ ) is 
 \be
 \begin{aligned}
q^\mu {\tilde J}^{abc}_{\mu\nu\lambda} (k_1,k_2)  =  &
 -\frac i{4\pi} f^{abc} \epsilon_{\lambda\nu \sigma}k_1^\sigma \,\frac
{2m}{k_1}\arccot\left( \frac {2m}{k_1} \right)\label{qJ4}\\
&- \frac i{4\pi} f^{abc} \epsilon_{\lambda\nu \sigma}k_2^\sigma \,\frac
{2m}{k_2}\arccot\left( \frac {2m}{k_2} \right).
\end{aligned}
\ee
Therefore, as far as the odd part is concerned, the 3-point conservation
(\ref{3ptconserv}) reads
\begin{multline}
 -i q^\mu \tilde J_{\mu\nu\lambda}^{(odd)abc}(k_1,k_2) +f^{abd} \tilde
J_{\nu\lambda}^{(odd)dc}(k_2)+ 
f^{acd} \tilde J_{\lambda\nu}^{(odd)db}(k_1)\label{qJ5}\\
 =- \frac 1{4\pi} f^{abc} \epsilon_{\lambda\nu \sigma}\left(k_1^\sigma \,\frac
{2m}{k_1}\arccot\left( \frac {2m}{k_1} \right)+ k_2^\sigma \,\frac
{2m}{k_2}\arccot\left( \frac {2m}{k_2} \right)\right)\\
+ \frac 1{4\pi} f^{abc} \epsilon_{\lambda\nu \sigma}\left(k_1^\sigma \,\frac
{2m}{k_1}\arccot\left( \frac {2m}{k_1} \right)+ k_2^\sigma \,\frac
{2m}{k_2}\arccot\left( \frac {2m}{k_2} \right)\right)=0.
\end{multline}
Thus conservation is secured for any value of the parameter $m$. The fact that
in the UV or IR limit we find a violation of the conservation is an artifact of the procedure
we have used  and we have to remedy by subtracting  suitable counterterms from the effective action. These subtractions are to be understood as (part of) the definition of our regularization procedure. 

The second remark concerns the odd-parity correlators we have obtained above in the IR limit, the 2-point
function $\sim \delta^{ab} \epsilon_{\mu\nu\sigma} k^\sigma$ and the 3-point
function $\sim f^{abc}  \epsilon_{\mu\nu\lambda}$. As expected from the fact
that they are correlators at a RG fixed point, both satisfy the Ward identities
of CFT, in particular the SCT one. They are both purely local and at least the 2-point one does not
come from the regularization of any bare correlator. Ref.\cite{giombi} provides
a classification of all bare correlators in 3d CFT, both odd- and even-parity ones.
These satisfy the simplest conservation law, in which lower order correlators are not involved.
It is clear that, a complete classification of CFT correlators requires that we add also 
those considered above, which satisfy the conservation law (\ref{3ptconserv}).

Another remark is that in many cases correlators can be constructed directly from free field theory
via the Wick theorem. It is evident that there is no conformal free field theory in 3d that can
give rise to the parity odd 2- and 3-point correlators found above.

Finally let us remark that similar results are expected in other odd dimensional
spacetimes. Interesting cases will be 7d for free fermions coupled to gravity,
and 5d and 7d for fermions coupled to a gauge field alone or to both gravity and
gauge fields.

\section{The Pontryagin trace anomaly}\label{s:Pontryagin}

The second example of a correlator made only of contact terms is in even
dimension, specifically in 4d. It is provided by the parity-odd 3-point function of
the energy-momentum tensor in which one of the entries is the trace of the
e.-m. tensor. This 3-point function is the basic (but not exclusive)
ingredient of the trace anomaly. It is well-known that in 4d a theory coupled to
external gravity is generically endowed with an energy-momentum tensor whose
trace takes the form
\be
T_\mu{}^\mu= a E+ c \EW^2 +e P,\label{emtrace}
\ee
where $E$ is the Euler density, $\EW^2$ the square Weyl density and $P$ the
density of the Pontryagin class
 \be
P=\frac 12\left(\frac{\epsilon^{nmlk}}{\sqrt{|g|}}R_{nmpq}R_{lk}{}^{pq}\right)\label{pontryagin}
\ee
where $\epsilon^{nmlk}$ is the numerical Levi-Civita symbol.
Our interest here focus on this term\footnote{Of course also the other anomalies, 
$E$ and ${\cal W}^2$, are local terms, but they come from the regularization of nonvanishing bare correlators.}. 
The obvious question is whether there are
models where this term  appears in the trace of the e.m. tensor, that is if
there are models in 4d where the coefficient $e$ does not vanish.  The natural
candidates are models involving chiral fermions, where the $\epsilon$ tensor may
appear in the trace of $\gamma$ matrices. The coefficient $e$ has been recently
calculated \cite{BGL,BDL}, following
an early work \cite{CD2}, (see also \cite{BGLBled13, Zagreb,Shapiro}) in a model
of free chiral fermions coupled to a background metric. 
 
The model is the simplest possible one: a right-handed spinor coupled to external
gravity in 4d. The action is
\be
S= \int d^4x \, \sqrt{|g|} \, i\bar \psi_R \gamma^m\left(\nabla_m +\frac 12 \omega_m
\right)\psi_R \label{action}
\ee
where $\gamma^m = e^m_a \gamma^a$, $\nabla$ ($m,n,...$ are world indices,
$a,b,...$ are flat indices) is the covariant derivative with respect to the
world indices and $\omega_m$ is the spin connection:
\be
\omega_m= \omega_m^{ab} \Sigma_{ab}\0
\ee
where $\Sigma_{ab} = \frac 14 [\gamma_a,\gamma_b]$ are the Lorentz generators.
Finally
$\psi_R= \frac {1+\gamma_5}2 \psi$. Classically the energy-momentum tensor 
\be
T_{\mu\nu}= \frac i2 \bar \psi_R \gamma_\mu
{\stackrel{\leftrightarrow}{\nabla}}_\nu\psi_R
 \label{emt}
\ee
is both conserved and traceless on shell. At one loop, to make sense of the
calculations one must introduce regulators. The latter generally break both
diffeomorphism and conformal invariance. A careful choice of the regularization
procedure may preserve diff invariance, but anyhow breaks conformal invariance,
so that the trace of the e.m. tensor takes the form (\ref{emtrace}), with
specific nonvanishing coefficients $a$, $c$ and $e$. There are various techniques to
calculate the latter: cutoff, point splitting, dimensional regularization, and a few others. 
Here, for simplicity we limit ourselves to a short summary of dimensional regularization. First one
expands the metric around a flat background:
$g_{\mu\nu}\approx\eta_{\mu\nu}+h_{\mu\nu}$, where $h_{\mu\nu}$ represent the
gravity fluctuation. Then one extracts from the action  propagator and vertices.
The essential ones are the fermion propagator  $\frac i{\slash \!\!\!
p+i\epsilon }$
and the two-fermion-one-graviton vertex ($V_{ffg}$) 
\be
-\frac i{8} \left[(p-p')_\mu \gamma_\nu + (p-p')_\nu \gamma_\mu\right] \frac
{1+\gamma_5}2\label{2f1g}
\ee
where $p,p'$ are the fermion momenta. The only contributing diagrams are the
triangle diagram together with the crossed one. The triangle diagram is
constructed by joining three vertices $V_{ffg}$ with three fermion lines. The
external momenta are $q$ (incoming) with labels $\sigma$ and $\tau$, and
$k_1,k_2$ (outgoing), with labels $\mu,\nu$ and $\mu',\nu'$ respectively. Of
course $q=k_1+k_2$. The internal momenta are $p $, $p-k_1$ and $p-k_1-k_2$,
respectively. After contracting $\sigma$ and $\tau$ the total contribution to
the 3-point e.m. tensor correlator, in which one of the entries is the trace,  is
\be 
&&-\frac 1 {256}\int \frac {d^4p}{(2\pi)^4}\, {\rm tr} \left[\left(\frac
1{\slashed{p}}\left((2p-k_1)_\mu \gamma_\nu+(\mu\leftrightarrow \nu)\right)
\frac 1{\slashed{p}-\slashed{k}_1}\right.\right.\label{T1}\\
&& \cdot\left((2p-2 k_1 - k_2)_{\mu'}\gamma_{\nu'}+(\mu'\leftrightarrow
\nu')\right) \left.\left.\frac 1{\slashed{p} - \slashed{k}_1 -\slashed{k}_2}
(2\slashed{p} -\slashed{k}_1 -\slashed{k}_2)\right) \frac {1+\gamma_5}2\right]\0
\ee
to which we have to add the cross diagram where $k_1,\mu,\nu$ is exchanged with
$k_2,\mu',\nu'$. This integral is divergent. To regularize it we use dimensional
regularization, which consists in introducing additional components of the
momentum running in the loop: $p\to p+l$, $l=(l_4,\ldots, l_{n-4})$. This
regulates the integral, and one can now proceed to the integration. Full details
of the calculation can be found in \cite{BGL,BDL}. The result is as follows.
Calling $\tilde T^{(tot)}_{\mu\nu\mu'\nu'}(k_1,k_2)$ the  overall contribution of
the two diagrams, with $k_1^2=k_2^2=0$, one has
\be
\tilde T^{(tot)}_{\mu\nu\mu'\nu'}(k_1,k_2)= \frac 1{3072\pi^2} \left(k_1\cdot
k_2  \,t_{\mu\nu\mu'\nu'\lambda\rho}\,- t^{(21)}_{\mu\nu\mu'\nu'\lambda\rho}
\right)
k_1^\lambda k_2^\rho\, \label{Ttot}
\ee
where
\be
t_{\mu\nu\mu'\nu'\kappa\lambda}&=&\eta_{\mu\mu'} \epsilon_{\nu\nu'\kappa\lambda}
+\eta_{\nu\nu'} \epsilon_{\mu\mu'\kappa\lambda} +\eta_{\mu\nu'}
\epsilon_{\nu\mu'\kappa\lambda} +\eta_{\nu\mu'}
\epsilon_{\mu\nu'\kappa\lambda},\0\\
t^{(21)}_{\mu\nu\mu'\nu'\kappa\lambda}&=&k_{2\mu}k_{1\mu'}
\epsilon_{\nu\nu'\kappa\lambda} +
k_{2\nu}k_{1\nu'}\epsilon_{\mu\mu'\kappa\lambda} +k_{2\mu}k_{1\nu'}
\epsilon_{\nu\mu'\kappa\lambda} +k_{2\nu}k_{1\mu'}
\epsilon_{\mu\nu'\kappa\lambda}.\0
\ee
Fourier transforming (\ref{Ttot}) and plugging the result in the full 1-point correlator of the e.m. tensor trace
\be 
\langle\!\langle T^{\mu}_{\mu}(x)\rangle\!\rangle= 2\sum_{n=1}^\infty \frac
{i^{n+1}}{(n-1)!} 
\int \prod_{i=2}^n dx_i\,  h_{\mu_i\nu_i}(x_i) \,\langle
0|\mathcal{T}\,T^{\mu}_{\mu}(x)\ldots T^{\mu_n\nu_n}(x_n)|0\rangle\label{Tmumu}
\ee
one obtains
\be 
\langle\!\langle T^{\mu}_{\mu}(x)\rangle\!\rangle = \frac
i{768\pi^2}\epsilon^{\mu\nu\lambda \rho} \left(\partial_\mu\partial_\sigma
h^\tau_\nu \, \partial_\lambda\partial_\tau
h_{\rho}^\sigma-\partial_\mu\partial_\sigma h^\tau_\nu \, 
\partial_\lambda\partial^\sigma h_{\tau\rho}\right)+\mathcal{O}(h^3),\label{final1}
\ee
which is the lowest order expansion in $h_{\mu\nu}$ of
\be 
\langle\!\langle T^{\mu}_{\mu}(x)\rangle\!\rangle = \frac i{768\pi^2} \, \frac
12\,\epsilon^{\mu\nu\lambda
\rho}R_{\mu\nu}{}^{\sigma\tau}R_{\lambda\rho\sigma\tau},\label{final2}
\ee
i.e. the Pontryagin trace anomaly. Changing chirality in (\ref{action}) leads
to a change of sign in the RHS of (\ref{final2}). Therefore, in left-right symmetric models
this anomaly is absent.  The surprising aspect of (\ref{final2}) is the
$i$ in the RHS. In other words the coefficient $e$ in (\ref{emtrace}) is
imaginary. Before entering the discussion of this point in the next section, 
let us recall that the odd-parity 3-point correlator, with three (untraced) e.m.
tensor insertions, in the model (\ref{action}), calculated by means of the Wick theorem,
identically vanishes in configuration space, \cite{BDL}. An unsurprising result, because on the basis
of a general theorem we know that the odd-parity conformal covariant 3-point e.m.
tensor bare correlator in 4d vanishes identically, \cite{Stanev1,Zhiboedov}. 

Finally let us remark that the one described in this section is not an isolated
case. Similar pure contact terms correlators (and similar anomalies) exist in $4k$
dimensions, and mixed gauge-gravity pure contact terms correlators may exist
also in other even dimensions. 

\subsection{Comments on the Pontryagin anomaly}\label{s:comments}

The Pontryagin anomaly is puzzling  at first because it looks like
a challenge for many commonplaces. Several points have been already discussed 
in section 4 of \cite{BGL} and in section 7 of \cite{BDL}. We would like to add here a few 
additional remarks. One surprising aspect of this anomaly is the appearance of an imaginary coefficient 
in front of it, with the consequence that the energy-momentum tensor at one loop becomes complex and may
endanger unitarity, see \cite{BGL}. The surprise is due to the fact that the action of the model 
(\ref{action}) is hermitean and one would not expect the e.m. tensor to become complex at one loop.
However this is a simple consequence of the regularization. For regularizing an expression may require
to trespass on the complex plane, much in the same way as when one looks for solutions of a real 
algebraic equation. The simplest example of this effect is the regularization of the real function 
$\frac 1x$ in one dimension given by ${\cal P}\frac 1x + \pi i \delta(x)$ 
(the first term is the principal value). Something similar happens in our regularization of (\ref{T1})
and leads to the imaginary coefficient of eq.(\ref{final2}). Therefore, finally, this result is not 
at all surprising.

An important aspect of the anomaly we are considering, which was only sketched in \cite{BDL}, is the following:
if instead of regularizing (\ref{T1}) (let's call it procedure ($a$)), as we have done above, 
we first regularize the 3-point function
of the untraced e.m. tensor and {\it then} take the trace of one of the insertions (procedure ($b$)), 
we get a vanishing result. It was pointed out in \cite{BDL} that the latter is not the correct way to
proceed. However, although this statement was supported by explicit examples in 2d, it may 
leave the impression that our result in \cite{BDL} and in the previous section is scheme dependent. 
This is not the case and we would like now to explain why. The point is that procedure ($b$), as just
outlined, is incomplete. As we have pointed out above regularizing may break not only Weyl 
symmetry but also diffeomorphism covariance. This is in fact what happens with both procedure ($a$)
and ($b$). But while, as was shown in \cite{BDL}, this breaking in case ($a$) is innocuous (one 
subtracts counterterms which restore covariance without modifying the trace anomaly), in case
($b$) the breaking of covariance is more substantial. In order to restore it one has to modify
the (previously vanishing) trace anomaly. The explicit calculation in scheme ($b$), which is very challenging, 
has not been done yet, but we conjecture that the result will restore 
the Pontryagin anomaly with the same coefficient as in (\ref{final2}). If this is true, as we believe,
choosing scheme ($a$) instead of ($b$) is only a matter of opportunity. 

We would like to add also a few words on a frequent source of misunderstanding, which stems from a 
reckless identification of Majorana and Weyl spinors in 4d. In 4d they transform according to two different 
irreducible representations of the Lorentz group. The first belong to a real representation and the 
second to a complex one. Moreover, Weyl fermions have definite chirality while for Majorana fermions 
chirality is not defined. Majorana fermion admit a massive term in the action, 
whereas Weyl fermions are rigorously massless. The corresponding Dirac operators are different, 
even in the massless case. So in no way can one confuse Majorana and Weyl spinors, 
even when massless. However misnaming is very frequent and not always innocuous, 
especially when anomalies are involved.

For instance, given a Weyl spinor $\chi$, one can construct a Majorana spinor $\psi$ as follows
\be 
\psi= \chi+ \gamma_0 C\chi^*\label{Majorana}
\ee
where $C$ is the charge conjugation matrix (for notation, see \cite{BGL}). 
If $\chi$ is left-handed, the conjugate spinor $\gamma_0 C\chi^*$ is right-handed. 
Thus we can see the reason why for Majorana fermions there is no Pontryagin anomaly.
But, apart from this, (\ref{Majorana}) is not much more than saying that the sum of a 
complex number and its conjugate is real. In any case it is not a good reason to
confuse Weyl and Majorana fermions. 

On the other hand many theories, in particular the supersymmetric ones, are conveniently 
formulated in terms of the two-component formalism, i.e. on the basis of
two-component spinors $\xi_\alpha$ and $\xi_{\dot\alpha}$ ($\alpha,\dot\alpha=1,2$). 
These two-component fields are the building blocks of the theory and, a priori, they
can be the components of either a Weyl, Majorana or Dirac fermion. When the two-component
formalism is used one must know the full content of the theory in order to decide 
that\footnote{Sometimes $\xi_\alpha$ and $\xi_{\dot\alpha}$ are called themselves Weyl spinors, which does not 
add to clarity.}. However the two-component formalism has many advantages, it serves well
for many purposes and there is no reason not to use it.  However the problem of
anomalies must be dealt with carefully, anomalies come from a (regularized) variation of 
the fermion determinant, i.e. the determinant 
of the relevant Dirac operator, which is different in the different cases. 
So when anomalies are involved it is of course irrelevant what formalism we use, provided we
unambiguously distinguish the true chiral nature of the fermions in the theory. 
For instance, it is a well known and important fact that consistent gravitational (Einstein) and Lorentz 
anomalies in 4d vanish. But this is not due to Weyl fermions being exchangeable
with Majorana ones, but rather because the 
third order symmetric invariant tensor of the Lorentz algebra vanishes identically. 
If one understand this it is not difficult to understand the origin of the Pontryagin anomaly.
In particular what is decisive for the latter is the overall balance of chirality.

\section{Conclusion}\label{s:conclusion}

Our purpose in this article was to show that in field theories, and in particular in conformal 
field theories, there are correlators made of pure contact terms, without a corresponding bare part.
We have exhibited two examples. The first obtained by considering the conformal limits of a free 
massive fermion theory in 3d and the current correlators thereof; we have shown 
that such correlators are in one-to-one correspondence with the terms of the 3d  gauge CS action.
The second corresponds to the case of the Pontryagin trace anomaly. Such an anomaly appears
in e.m. tensor correlators containing one trace insertion. We have shown that
the corresponding 3-point correlator is nonvanishing even though the corresponding untraced 
correlator vanishes (that is, there is no bare correlator underlying it).
In other words pure contact term correlators may live of their own.

\section*{Acknowledgements} One of us (L.B.) would like to thank the organizers of the 18th Workshop 
"What Comes Beyond the Standard Models", Bled, July 11-19, 2015, for giving him the opportunity 
to present the previous material.


\end{document}